\documentclass[preprint,showpacs]{revtex4-1}
\usepackage{graphicx}
\usepackage{xcolor}
\usepackage{amssymb}
\newcommand{\be}{\begin{eqnarray}}
\newcommand{\ee}{\end{eqnarray}}
\newcommand{\nn}{\nonumber}

\newcommand{\ms}{\mskip 1.5mu}

\begin{document}

\title{Double parton distributions for a positronium-like bound state using light-front wave functions}
\author{\bf Chandan Mondal$^a$, Asmita Mukherjee$^b$ and Sreeraj Nair$^a$}

\affiliation{$^a$ Institute of Modern Physics, Chinese Academy of Sciences, Lanzhou 730000, China \\$^b$
 Department of Physics,
Indian Institute of Technology Bombay,Powai, Mumbai 400076,
India.}

\begin{abstract}
We investigate the double parton distributions (DPDs) for a positronium-like bound state using light-front QED. We incorporate the higher Fock three particle component of the state, that includes a photon. We obtain the overlap representation of the DPDs in terms of the three-particle light-front wave functions (LFWFs). Our calculation explores the 
correlations between the momentum fractions of the particles  probed and the transverse distance between them, without any assumption of factorization between them. We also investigate the behavior of the DPDs near the kinematical boundary when the sum of the momentum fractions is close to one.   

\end{abstract}
\maketitle
\section{ Introduction}
As the flux of partons increases in high energy hadronic collision experiments, the probability of having more than one independent hard scattering interaction also increases and a proper description of final states in hadronic collisions requires the inclusion of multiple partonic interactions (MPIs). The MPIs in hadronic collision have been predicted a long ago \cite{old_collection1,old_collection2,old_collection3,old_collection4,old_collection5}. 
The most probable and the simplest of these MPIs are the double parton scattering (DPS) events. In DPS two partons from each hadron participate in separate hard interactions. In such a process, a large momentum transfer is
involved in both scattering. The first experimental evidence of DPS was found at CERN-ISR \cite{cern_isr} in p-p collision. DPS are indeed relevant at LHC because of the high density of partons. The ATLAS collaboration had reported their first results on DPS a while ago \cite{atlas_dpd} and DPS also contributes to the Higgs production background in several channels at LHC. 

DPS can be factorized in terms of the hard interactions which are calculable in perturbation theory and the double parton distribution  (DPDs) functions. The DPDs depend on two-body quantities encoding the non-perturbative dynamics of the partons. Factorization of DPS usually assumes the simplest case wherein there are no correlations between the two partons \cite{Diehl1,manohar1,Diehl2,Diehl3}. The DPDs
are interpreted as the number densities of  parton pair at a given  transverse distance  $y_\perp$ and carrying  longitudinal momentum fractions 
($ x_1,x_2$) of the composite system \cite{old_collection2,Diehl2,Calucci:1999yz}. Since the DPDs 
depend on the partonic inter-distance \cite{Calucci:1999yz},
they contain information on the hadronic structure    
which compliments the tomographical information encoded by the one-body distributions such as generalized parton distributions (GPDs) \cite{gpd} and transverse momentum dependent distributions (TMDs) \cite{tmd}. Therefore, DPDs represent a novel tool to access the three-dimensional hadron structure. Despite the wealth of information provided by the DPDs, the present experimental knowledge is mainly accessible through the 
DPS cross section which has been accumulated into the
effective cross section, $\sigma_{\mathrm{eff}}$. For the recent results we refer to the articles \cite{data6,data8,data9,data10,data11,data12}.  

DPDs being nonperturbative in nature are always very difficult to evaluate from QCD first principles and there have been numerous attempts to gain insight into them by studying QCD inspired models. Model calculations of DPDs are important and interesting to understand the properties as well as for predictions of experimental observables.  
Several phenomenological models such as bag model \cite{Mel_19}, constituent quark model \cite{cqm1,cqm2,cqm3,sm}, generalized valon model \cite{Mel_23}, soft-wall AdS/QCD model \cite{ads1}, dressed quark model \cite{Kasemets:2016nio} etc. have been used to obtain the basic information on DPDs and to gauge the phenomenological impact of transverse and longitudinal correlations, along with spin correlations  \cite{cqm2,cotogno,muld,positivity,Kasemets:2012pr}.
The transverse  structure of of the proton from the DPDs and the effective cross section has been investigated in \cite{Calucci:1999yz,Rinaldi:2018bsf,Rinaldi:2015cya}. Recently, the matching of both the position and momentum space DPDs onto ordinary parton distribution functions at the next-to-leading order (NLO) in perturbation theory has been reported in \cite{Diehl:2019rdh}, where the authors have also discussed about the  sum rules for DPDs \cite{Diehl:2018kgr}.
The quantities related to DPDs, and encoding double parton correlations, have been evaluated for the pion in lattice QCD \cite{lattice}.

As very little is known so far on the DPDs $F(x_1,x_2, y_\perp)$, there are  several approaches to parameterize or model them.  A common approach is based on a factorized  ansatz, which assumes that the  $y_\perp$ dependence is factored out from the dependence on $x_1$ and $x_2$, which are the momentum fractions of the partons probed. In addition, it is sometimes also assumed that $x_1$ and $x_2$ dependences are factored out, in terms of single parton distributions (pdfs) and neglecting any correlations between them \cite{Mel_19}. In \cite{Rinaldi:2016jvu} an approach was used based on \cite{Diehl:2014vaa} where  the DPDs were written as a convolution of two impact parameter dependent pdfs which are obtained from GPDs. A Gaussian form of the impact parameter dependent pdfs were used. However it was concluded that the factorized ansatz fails in the valence region and the authors also observed that a Gaussian dependence on $y_\perp$ is rather arbitrary. It is thus relevant to investigate the DPDs without such assumptions. The model calculations  can be thought of as a parameterization of the DPDs at a   low momentum scale and one then evolves them to a higher scale of the experiments using evolution equation, such evolution equations have been obtained by now and discussed in detail \cite{Gaunt:2009re}. Another interesting aspect of model calculation of the DPDs is the behavior near the kinematical bound $x_1+x_2 =1$. The DPDs should vanish in the unphysical region $x_1+x_2 > 1$. In some early model calculations this support property was violated, due to non-conservation of momentum of the constituents. In later calculations, a phenomenological factor is included to improve the behavior in this kinematical limit. 

A widely used method  to calculate the DPDs is by expressing them in terms of overlaps of light-front wave functions (LFWFs). In the light-front formalism, the proton state is expanded in Fock space in terms of multi-parton LFWFs. The LFWFs satisfy the bound state equation in light-front(LF) QCD. One then truncates the Fock space to a few particle sector; such truncation is boost invariant in light-front framework. As it is very difficult to solve the light-front bound state equation, in particular to obtain the wave functions of the higher Fock sector, most model calculations are restricted to using the three
quark valence LFWF for the proton. In a previous work \cite{Kasemets:2016nio}, in order to calculate the quark-gluon DPDs, a different approach was used; namely instead of a proton state a relativistic spin 1/2 composite state of a quark dressed with a gluon was used. The LFWFs of the two-particle state was calculated in perturbation theory. This may be thought of as a field theory based perturbative model, to investigate the quark-gluon correlations in the DPDs. However, the kinematics of a two-particle system are rather constrained. In this work, we use the overlap approach in terms of LFWFs for a two-particle bound state like a positronium in QED, in the weak coupling limit.  We include the effect of the three particle $e^+ e^- \gamma$ component of the LFWF.  As solving the LF bound state equation is rather difficult in QED as well, we use a simpler but nevertheless interesting approach earlier followed in \cite{Mukherjee:2001ra}, to calculate the twist-four structure function of positronium and verifying a sum rule.  We use an  analytic form of the  two-particle LFWF in the weak coupling limit. The three-particle LFWF is then expressed in terms of the two-particle LFWF using LF QED Hamiltonian. This calculation illustrates the formalism which can also be applied to a QCD mesonic system; in fact in the weak coupling limit, the LFWFs are expected to mimic those  of a meson. Our approach allows us to calculate them without any assumption on factorization of the $x_1, x_2$ and $y_\perp$ dependence, and we can investigate the interplay between these variables in full form. Thus, our calculation may be thought of as an exploratory analysis on the explicit $x_1,x_2$ and $y^\perp$ dependence of the DPDs in a three-particle system. We also discuss the behavior of the DPD in the limit $x_1+x_2 \to 1$.

The paper is organized as follows: In section~\ref{overlap} we discuss the DPDs for the $\{e^{-}e^{+}\}$ pair and their overlap representation in the light-front dressed positronium model. We present the numerical results in section~\ref{results}. Conclusions are given in section~\ref{conclusion}.
\section{Double Parton Distributions }\label{overlap}

The double parton distributions (DPDs) for unpolarized quarks  can be defined as \cite{Diehl2,Diehl3}
\be
 F_{a_1a_2}(x_1,x_2,{\bf y^{\perp}})
  &=& 2p^+ 
        \int \frac{dz^-_1}{2\pi}\, \frac{dz^-_2}{2\pi}\, dy^-\;
           e^{i\ms ( x_1^{} z_1^- + x_2^{} z_2^-)\ms p^+}\nonumber\\
&\times&\left<p|\, \mathcal{O}_{a_2}(0,z_2)\,
            \mathcal{O}_{a_1}(y,z_1) \,|p\right> \,,
\ee

\noindent
where $\mid p\rangle $ is the target system with momentum $p$. $x_1$ and $x_2$ are momentum fractions 
of the partons and ${\bf y^{\perp}}$ is the relative transverse distance between them. 

The fermionic operators are given by \cite{Diehl2} (see the appendix)
\be
\mathcal{O}_{a_i}(y,z_i) = \bar{\psi_i}\left(y-\frac{z_i}{2}\right)\Gamma_{a_i}
\psi_i\left(y-\frac{z_i}{2}\right)\Big{|}_{z_i^+=y_i^+=0,z_i^{\perp}=0}
\label{op}\ee
\noindent
where $\Gamma_{a_i}$ are various Dirac $\gamma$ matrices projecting onto the corresponding polarization 
states given by
\be
\Gamma_{q} = \frac{1}{2}\gamma^+ , \hspace{1cm} \Gamma_{\Delta q} = \frac{1}{2}\gamma^+ \gamma_5,
\hspace{1cm} \Gamma^j_{\delta q } = \frac{1}{2}i \sigma^{j+}\gamma_5
\ee
\noindent
for the unpolarized fermion $(q)$, longitudinally polarized fermion $(\Delta q)$ or transversely polarized
fermion $(\delta q)$ respectively. We choose the light-cone gauge and the gauge link in the operator structure is set to unity.
\subsection{Overlap Representation for the DPD}
As discussed in the Introduction, we consider our target state to be a positronium-like bound state in LF QED. We use the two-component form of the LF QED in the line of \cite{Brodsky:2006ku, hari}.  In this section, we present a calculation of the unpolarized fermion DPDs for such a state.
This means $\Gamma_{q} = \frac{1}{2}\gamma^+$ in Eq.~\ref{op}.
The state can be expanded in Fock space in terms of LFWFs as
\be
\mid P \rangle = && \sum_{\sigma_1, \sigma_2} 
\int {dp_1^+ d^2 p_1^\perp \over \sqrt{2 (2 \pi)^3 p_1^+}} 
\int {dp_2^+ d^2 p_2^\perp \over \sqrt{2 (2 \pi)^3 p_2^+}} 
\nonumber \\
&& \phi^2(P \mid p_1, \sigma_1; p_2, \sigma_2) \sqrt{2 ((2 \pi)^3 P^+}
\delta^3(P-p_1-p_2) b^\dagger(p_1, \sigma_1) d^\dagger(p_2,\sigma_2) \mid 0
\rangle \nonumber \\
&&~~~~~~ + \sum_{\sigma_1,\sigma_2,\lambda} 
\int {dp_1^+ d^2 p_1^\perp \over \sqrt{2 (2 \pi)^3 p_1^+}} 
\int {dp_2^+ d^2 p_2^\perp \over \sqrt{2 (2 \pi)^3 p_2^+}} 
\int {dp_3^+ d^2 p_3^\perp \over \sqrt{2 (2 \pi)^3 p_3^+}} 
\nonumber \\
&&~~~~~~~~ \phi^3(P \mid p_1, \sigma_1; p_2, \sigma_2; p_3, \lambda)
\sqrt{2 (2 \pi)^3 P^+} \delta^3(P-p_1 -p_2 -p_3) 
\nonumber \\
&&~~~~~~~~~~~~~~~~~~b^\dagger(p_1 ,\sigma_1)
d^\dagger (p_2, \sigma_2) a^\dagger(p_3, \lambda) \mid 0 \rangle, 
\label{target}
\ee
\noindent
where the first term corresponds the two particle Fock sector, $\mid e^+e^-~\rangle$ with the two particle LFWF $\phi^2$ and the second term is the three particle Fock component $\mid e^+e^-\gamma ~\rangle$ wherein $\phi^3$ is the three particle LFWF. $\sigma_1$, $\sigma_2$ and $\lambda$ are the helicities of the electron, positron and photon respectively.
The LFWF are written in terms of the Jacobi momenta ($x_i,q_{i}^\perp$) defined as
\be
p_i^+ = x_i p^+,  \hspace{1cm} p_{i}^{\perp} = q_{i}^{\perp} +x_i p^{\perp}
\ee
\noindent
where $\sum_i x_i =1$ and $\sum_i q_{i}^{\perp} =0$.
The contribution coming from the three particle sector of the Fock space can then be written in term of overlap of LFWFs,
\be\label{dpdeq}
F_{e^-e^+}(x_1,x_2,{\bf y^{\perp}})
&=&\frac{(p^+)^2}{2\pi^2} ~\sum_{\sigma_1,\sigma_1^\prime,\sigma_2,\sigma_2^\prime,\lambda} ~\int d^2k_1^{\perp}~d^2k_2^{\perp}~d^2k_1'^{\perp}~
\phi^{3*}_{\sigma_1,-{\sigma_2^\prime},\lambda}(p,k_1,k_1',p-k_1-k_1')\nonumber\\
&&\phi^{3}_{{\sigma_1^\prime},-\sigma_2,\lambda}(p,k_1+k_1'-k_2,p-k_1-k_1')~e^{i(k_1^\perp-k_1'^\perp).y^\perp}
\ee
with
$
p^+\phi^3_{\sigma_1\sigma_2\lambda}(k_i^+,k_i^\perp)=\psi^3_{\sigma_1\sigma_2\lambda}(x_i,q_i^\perp).
$
The Eq.(\ref{dpdeq}) can be rewritten as
\be
&&F_{e^-e^+}(x_1,x_2,{\bf y^{\perp}})\nonumber\\
&=&\frac{1}{2\pi^2} ~\sum_{\sigma_1,\sigma_1^\prime,\sigma_2,\sigma_2^\prime,\lambda} \int d^2k_1^{\perp}~d^2k_2^{\perp}~d^2{{k'}}_1^{\perp}~
\psi^{3*}_{\sigma_1,-{\sigma_2^\prime},\lambda}(x_1,k_1^\perp;x_2,k_1'^\perp + k_2^\perp-k_1^\perp;1-x_1-x_2,k_3^\perp)\nonumber\\
&\times&\psi^{3}_{{\sigma_1^\prime},-\sigma_2,\lambda}(x_1,{{k'}}_1^\perp;x_2,k_2^\perp;1-x_1-x_2,k_3^\perp)~e^{i(k_1^\perp-{{k'}}_1^\perp).y^\perp}\label{DPD_P}
\ee
where $k_3^\perp=p^\perp-k_1'^\perp-k_2^\perp$, we consider the frame where $p^\perp=0$.
The amplitudes or LFWFs $\psi^2$ and  $\psi^3$ are boost invariant and are functions of the Jacobi momenta. These  can be written as \cite{Harindranath:1998pd,hari},
\begin{eqnarray}\label{psi3}
\psi^3_{\sigma_1, \sigma_2, \lambda_3}(x_1, k_1; x_2, k_2;
1-x_1-x_2, k_3) = {\cal M}_1 + {\cal M}_2,
\end{eqnarray}
where the amplitudes are given by \cite{Mukherjee:2001ra} :
\begin{eqnarray}
	{\cal M}_1 &=& { 1 \over E} (-) { e \over \sqrt{2 (2 \pi)^3}}
		{ 1 \over \sqrt{1 - x_1 - x_2}} ~V_1~ \psi^2_{s_1, 
		\sigma_2}(1-x_2, -k_2^\perp; x_2,k_2^\perp), \nonumber\\
	{\cal M}_2 &=& { 1 \over E} { e \over \sqrt{2 (2 \pi)^3}}
		{ 1 \over \sqrt{1 - x_1 - x_2}} ~V_2~ \psi^2_{\sigma_1, 
		s_2}(x_1,k_1^\perp;1-x_1,-k_1^\perp) 
\end{eqnarray}
with the energy denominator
\begin{eqnarray}
	E(x_1,x_2)= \Big[ M^2  - {m^2 + (k_1^\perp)^2 \over x_1} -
		{m^2 + (k_2^\perp)^2 \over x_2} - {(k_3^\perp)^2 
		\over 1 - x_1 -x_2} \Big ],\nonumber
\end{eqnarray}  
and the vertices     
\begin{eqnarray}\label{v}
	V_1(x_1,k_1^\perp;x_2,k_2^\perp)&=&\chi_{\sigma_1}^\dagger \sum_{s_1}\left [ { 2 
	k_3^\perp \over 1 - x_1 -x_2} - { (\sigma^\perp. k_1^\perp
	- i m) \over x} \sigma^\perp + \sigma^\perp {(\sigma^\perp. 
	k_2^\perp -im) \over 1-x_2} \right] \chi_{s_1}. 
		(\epsilon^\perp_{\lambda_1})^*,\nonumber\\
	V_2(x_1,k_1^\perp;x_2,k_2^\perp)&=&\chi_{-\sigma_2}^\dagger \sum_{s_2}
		\left [ { 2 k_3^\perp \over 1 - x_1 -x_2} - \sigma^\perp
		{ (\sigma^\perp. k_2^\perp
		- i m) \over x_2}  +  {(\sigma^\perp. k_1^\perp -
		im) \over 1-x_1} \sigma^\perp 
		\right] \chi_{-s_2}. (\epsilon^\perp_{\lambda_1})^*.\nonumber\\
\end{eqnarray}
The above expressions are obtained using the light-front Hamiltonian for QED in a similar line as in light-front QCD 
\cite{hari}. Following Eqs.($\ref{psi3}-\ref{v}$) we can rewrite the Eq.($\ref{DPD_P}$) in terms of $\psi^2$ as 

\be
F_{e^-e^+}(x_1,x_2,{\bf y^{\perp}})
&=&{e^2 \over {(2 \pi)^5}}\frac{1}{[E(x_1,x_2)]^2} \frac{1}{1-x_1-x_2}  ~\sum_{\sigma_1,\sigma_1^\prime,\sigma_2,\sigma_2^\prime,\lambda} \int d^2k_1^{\perp}~d^2k_2^{\perp}~d^2{{k'}}_1^{\perp}~
\nonumber\\
&\times&[\mathcal{P}_{11}+\mathcal{P}_{12}+\mathcal{P}_{21}+\mathcal{P}_{22}]~e^{i(k_1^\perp-{{k'}}_1^\perp).y^\perp}\label{mainres}
\ee
where
\be
\mathcal{P}_{11}&=&[V_1(x_1,k_1^\perp;x_2,{k'}_1^\perp +k_2^\perp - k_1^\perp)~\psi^2_{s_1, 
		-\sigma_2^\prime}(1-x_2, -{k'}_1^\perp -k_2^\perp + k_1^\perp; x_2,{k'}_1^\perp +k_2^\perp - k_1^\perp)]^{\dagger}\nonumber\\
&\times&[V_1(x_1,{{k'}}_1^\perp;x_2,k_2^\perp)~\psi^2_{s_1, 
		-\sigma_2}(1-x_2, -k_2^\perp; x_2,k_2^\perp)]\nonumber\\
\mathcal{P}_{22}&=&[V_2(x_1,k_1^\perp;x_2,{k'}_1^\perp +k_2^\perp - k_1^\perp)~\psi^2_{\sigma_1, 
		s_2}(x_1, k_1^\perp; 1-x_1,-k_1^\perp)]^{\dagger}\nonumber\\
&\times&[V_2(x_1,{{k'}}_1^\perp;x_2,k_2^\perp)~\psi^2_{\sigma_1^\prime,
		s_2}(x_1,{{k'}}_1^\perp; 1-x_1,-{{k'}}_1^\perp)]	\nonumber\\
\mathcal{P}_{12}&=&[V_1(x_1,k_1^\perp;x_2,{k'}_1^\perp +k_2^\perp - k_1^\perp)~\psi^2_{s_1, 
		-\sigma_2^\prime}(1-x_2, -{k'}_1^\perp -k_2^\perp + k_1^\perp; x_2,{k'}_1^\perp +k_2^\perp - k_1^\perp)]^{\dagger}\nonumber\\
&\times&[V_2(x_1,{{k'}}_1^\perp;x_2,k_2^\perp)~\psi^2_{\sigma_1^\prime,
		s_2}(x_1,{{k'}}_1^\perp; 1-x_1,-{{k'}}_1^\perp)]\nonumber\\
\mathcal{P}_{21}&=&[V_2(x_1,k_1^\perp;x_2,{k'}_1^\perp +k_2^\perp - k_1^\perp)~\psi^2_{\sigma_1, 
		s_2}(x_1, k_1^\perp; 1-x_1,-k_1^\perp)]^{\dagger}\nonumber\\
&\times&[V_1(x_1,{{k'}}_1^\perp;x_2,k_2^\perp)~\psi^2_{s_1, 
		-\sigma_2}(1-x_2, -k_2^\perp; x_2,k_2^\perp)].
\ee

The above expression is evaluated using Mathematica to calculate the spinor products. The final expressions are given below. 

\be
\mathcal{P}_{11}&=& \frac{8 (1+x_1-x_2)^2 \Big(k_1'^y (1-x_2) + k_2^y x_1 \Big) \Big( 
(k_1'^y +k_2^y) x_1 - k_1^y (x_1+x_2-1)
\Big)}{x_1^2 (x_2-1)^2 (x_1+x_2-1)^2} \times \nn \\
&&~~~~~~~~~~~~~~~~~~~~~~~~~~~~~~~~~~~~~~~~~~~~~~~~~~~
\psi^2\left(x_2,k_1^{\perp} - k_1'^{\perp} - k_2^{\perp}\right) \psi^2(x_2,k_2^{\perp})
\ee

\be
\mathcal{P}_{22}&=& \frac{-8 (1-x_1+x_2)^2 \Big(k_2^y (x_1-1) + k_1'^y x_2 \Big) \Big( 
 k_1^y (x_1+x_2-1) -(k_1'^y +k_2^y) (x_1-1) 
\Big)}{x_2^2 (x_1-1)^2 (x_1+x_2-1)^2} \times \nn \\
&&~~~~~~~~~~~~~~~~~~~~~~~~~~~~~~~~~~~~~~~~~~~~~~~~~~~~~~
\psi^2\left(x_1,k_1^{\perp}\right) \psi^2(x_1,k_1'^{\perp})
\ee

\be
\mathcal{P}_{12}&=& \frac{8((x_1-x_2)^2-1)\Big(k_2^y (1-x_1) + k_1'^y x_2 \Big)\Big(
k_1^y (x_1+x_2-1) - (k_1'^y + k_2^y) x_1
\Big)
}{x_1 (x_1-1) x_2 (x_2-1) (x_1+x_2-1)^2} \times \nn \\
&&~~~~~~~~~~~~~~~~~~~~~~~~~~~~~~~~~~~~~~~~~~~~~~
\psi^2\left(x_2,k_1^{\perp} - k_1'^{\perp} - k_2^{\perp}\right) \psi^2(x_1,k_1'^{\perp})
\ee

\be
\mathcal{P}_{21}&=& \frac{8((x_1-x_2)^2-1)\Big(k_1'^y (1-x_2) + k_2^y x_1 \Big)\Big(
  (k_1'^y + k_2^y) (x_1 -1) - k_1^y (x_1+x_2-1)
\Big)
}{x_1 (x_1-1) x_2 (x_2-1) (x_1+x_2-1)^2} \times \nn \\
&&~~~~~~~~~~~~~~~~~~~~~~~~~~~~~~~~~~~~~~~~~~~~~~~~~~~~~~~~~~~~~~~~~
\psi^2\left(x_1,k_1^{\perp}\right) \psi^2(x_2,k_2^{\perp})
\ee

Motivated by \cite{jones, Mukherjee:2001ra}, we take the two-particle wave-function $\psi^2$ in the weak coupling limit as :
\be
\psi^2(x,k^{\perp}) = \sqrt{\frac{m}{\pi^2}}  \frac{4(e_1)^{5/2}}{\left[  
(e_1)^2 -m^2 +\frac{1}{4}\frac{(k^{\perp})^2+m^2}{x(1-x)}
\right]^2},
\ee
with $m$ is the electron mass and $e_1=m/2$.
\begin{figure}[!htb]
\centering
\includegraphics[width=8.0cm,height=6cm,clip]{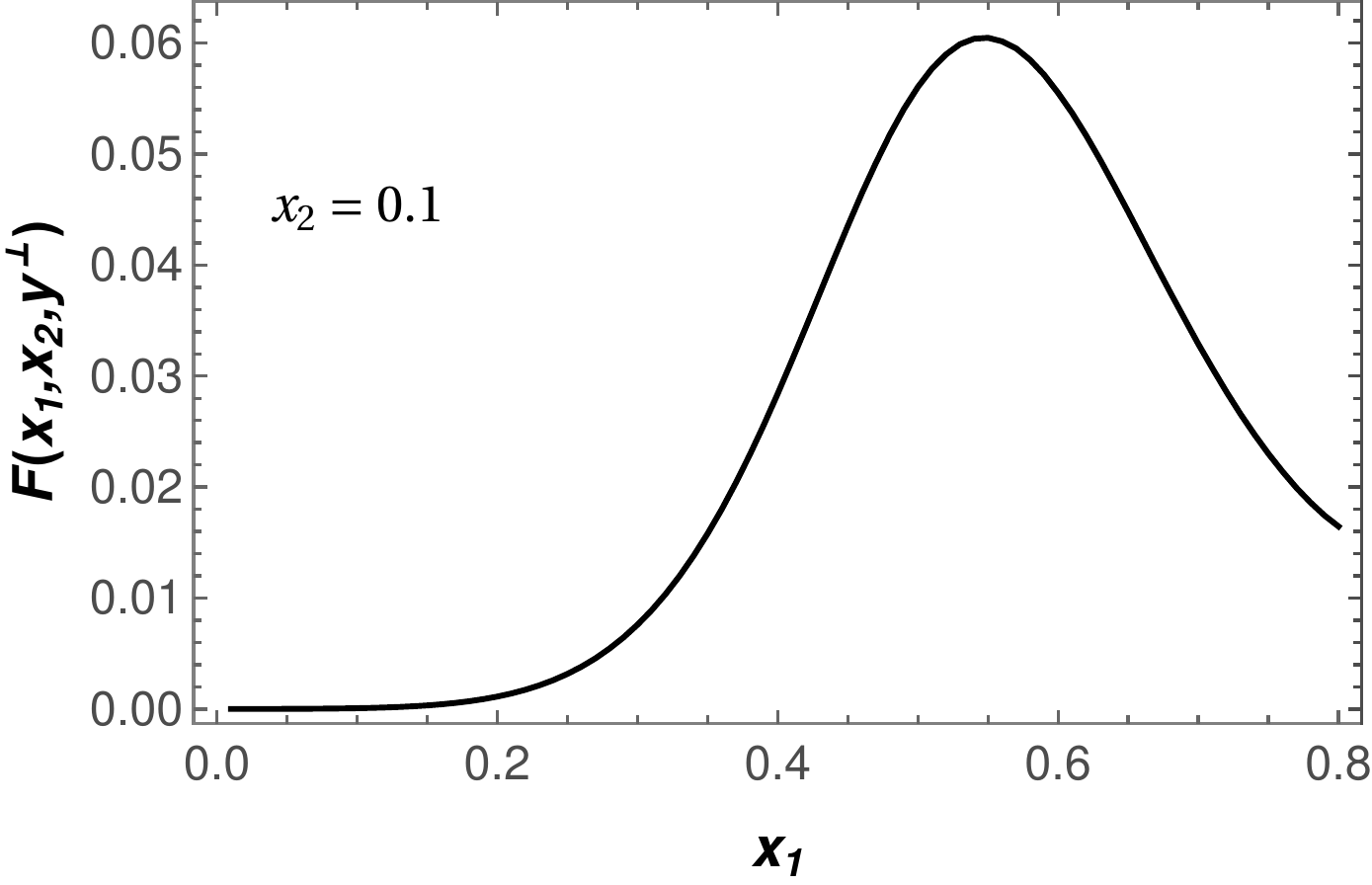}
\includegraphics[width=8.0cm,height=6cm,clip]{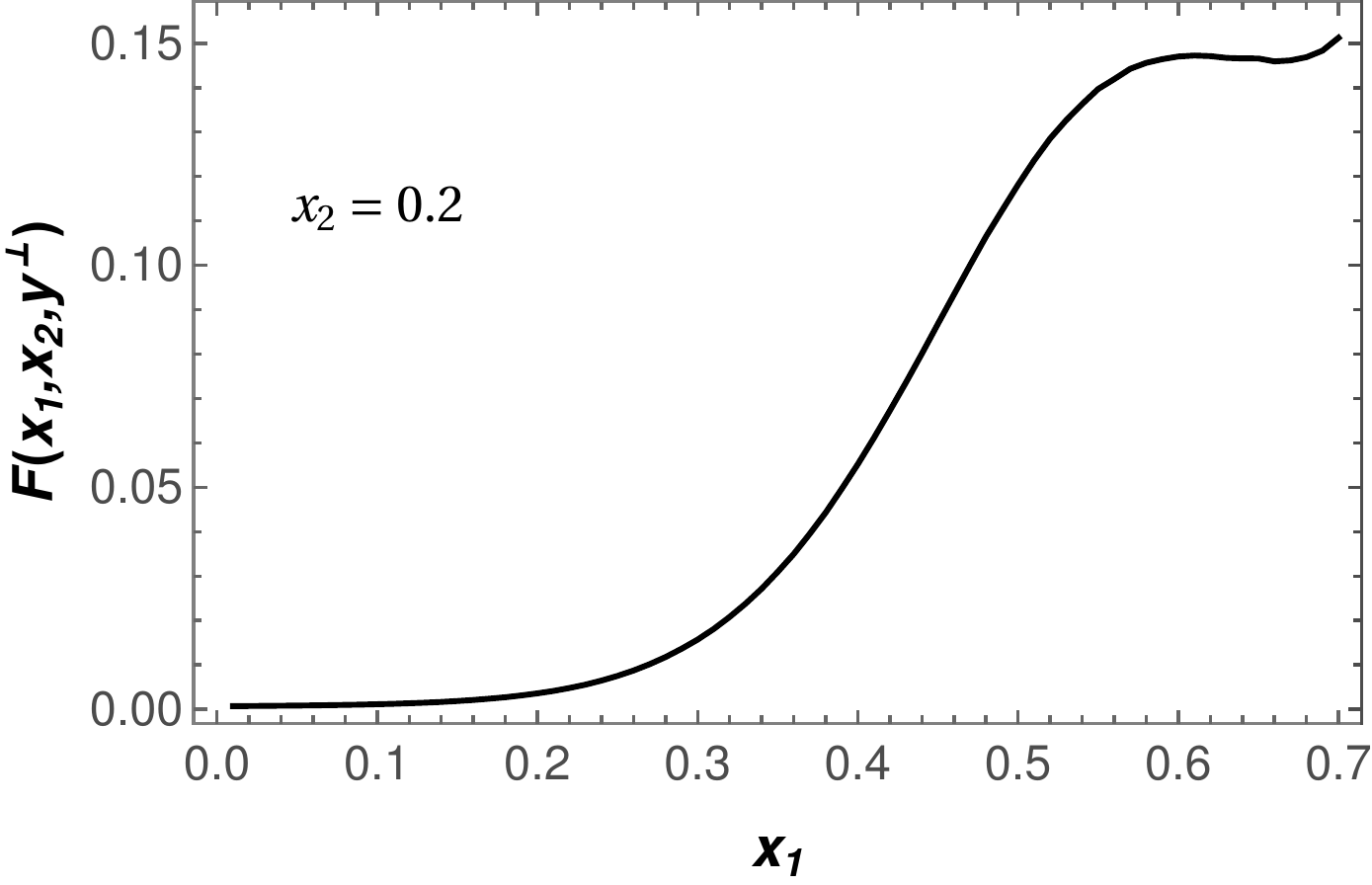}
\includegraphics[width=8.0cm,height=6cm,clip]{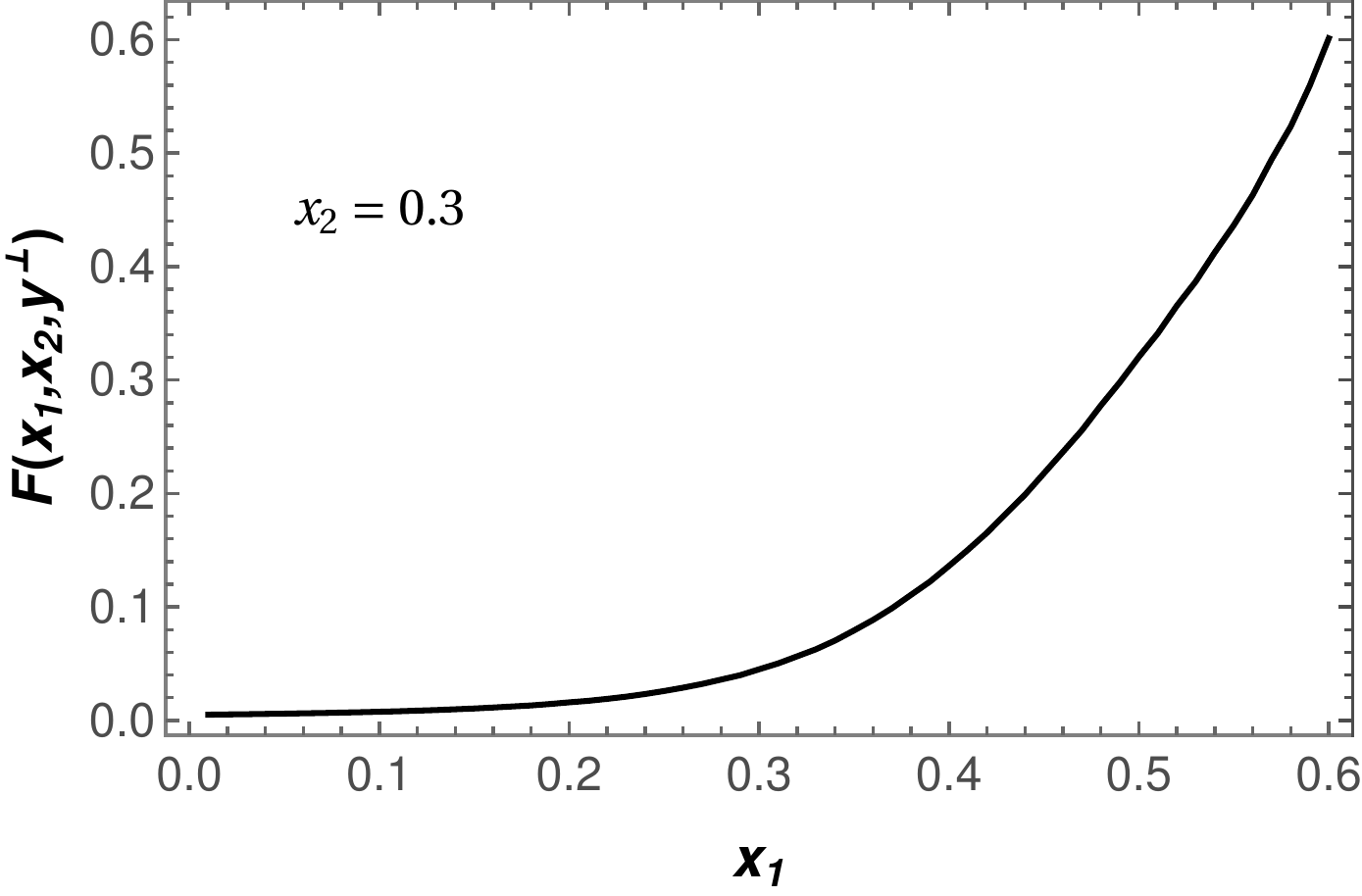}
\caption{Plot for $F(x_1,x_2,y^{\perp})$ vs $x_1$ for fixed value of 
$y^{\perp} = 0.2$ }
\label{fig1}
\end{figure}

\section{Numerical Results}\label{results}
In this section we present the numerical results for the unpolarized DPDs for the correlation between $e^-$ and $e^+$. In order to do the numerical calculations, a cut-off $k_{\rm max}=20$ MeV has been introduced for the upper limit of all the  integrations over $k^{\perp}$. We notice that for higher vales of $k_{\rm max} $ the results do not change. The electron (positron) mass has been taken as $m=0.50~\mathrm{MeV}$.  
In Fig~\ref{fig1}, we show the DPD for $e^{-}$ and $e^{+}$ pair in the positronium-like  bound state as a function of $x_1$ for different values of $x_2$ and fixed value of $y^{\perp}= 0.2$ MeV$^{-1}$. In this figure, we present the contribution evaluated from the $|e^-e^+\gamma\rangle$ Fock sector.  In our calculation, we have chosen the physical kinematical region in the three-particle sector, that is,  $x_1+x_2 < 1$. We observe that as $x_1 + x_2 \rightarrow 1$, the distribution shows a sharp rise. This is expected since the analytic expression from the three-particle section has a pole at $x_1+x_2=1$.  Actually the two-particle sector contributes at $x_1+x_2=1$, in particular to the normalization of the state, and it is necessary to incorporate this contribution to correctly predict the behavior of the DPDs at $x_1+x_2=1$, similar to the calculation of single parton distribution, or the structure functions \cite{Harindranath:1998pd}. In the calculation of the pdfs, the normalization contribution coming from the state cancels the pole.  It is beyond the scope of the present work to investigate if such cancellation happens for the DPDs, rather, we follow a more phenomenological approach, as seen later in this section.  It can also be noticed that there is a peak in the distribution near $x_1 \approx 0.5 $ for lower value of $x_2$. However, the peak disappears as the magnitude of the distributions increase significantly with increasing $x_2$ and the distribution behaves like ordinary parton distribution function of the bare electron in a physical electron system \cite{Chakrabarti:2014cwa}. In Fig~\ref{fig1} the peak is present only for $x_2=0.1$ because of the term $\left(x_1-1\right){}^2 x_2^2 \left(x_1+x_2-1\right){}^2$ present in the denominator of the term $\mathcal{P}_{22}$ which suppress the peak value for $x_2 > 0.1$.  

\begin{figure}[!htb]
\centering
\includegraphics[width=8cm,height=6cm,clip]{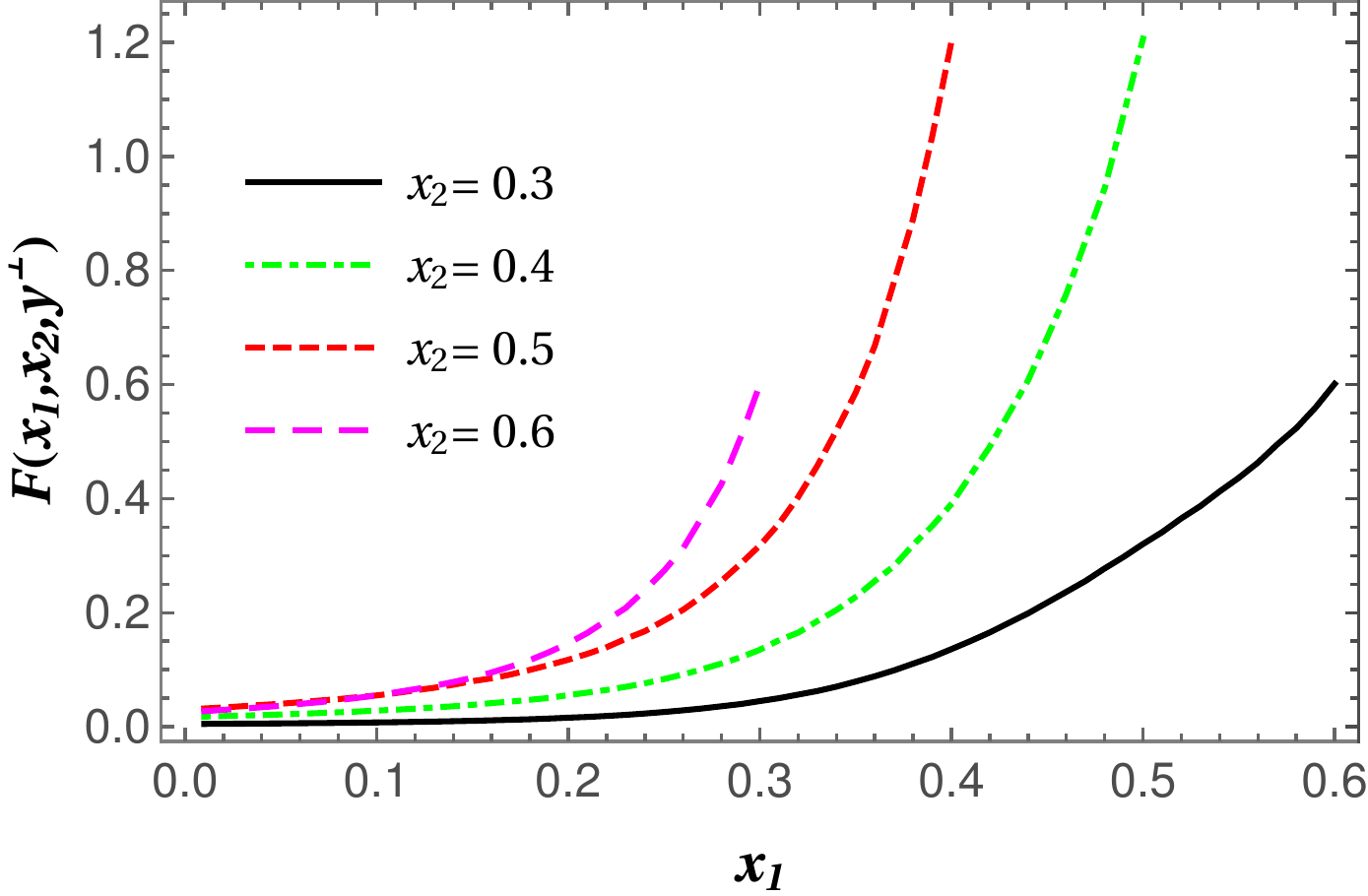}
\caption{(Color online) 3D plot for $F(x_1,x_2,y^{\perp})$ vs $x_1$ and $y^{\perp}$ for fixed value of 
$x_2 = 0.3$ }\label{fig2}
\end{figure}

The DPD as a function of $x_1$ for four different values of $x_2$ together with fixed $y^{\perp}=0.2$ MeV$^{-1}$ is shown in Fig~\ref{fig2}. We notice that the maximum value for the distribution is obtained near $x_1=0.5$ when $x_2=0.5$, and the magnitude of the distribution gradually decreases on both sides of $x_2=0.5$. This implies that the correlation of the $e^-e^+$ pair is maximum when they share the total momentum of the system equally.

\begin{figure}[!htb]
\centering
\includegraphics[width=8cm,height=6cm,clip]{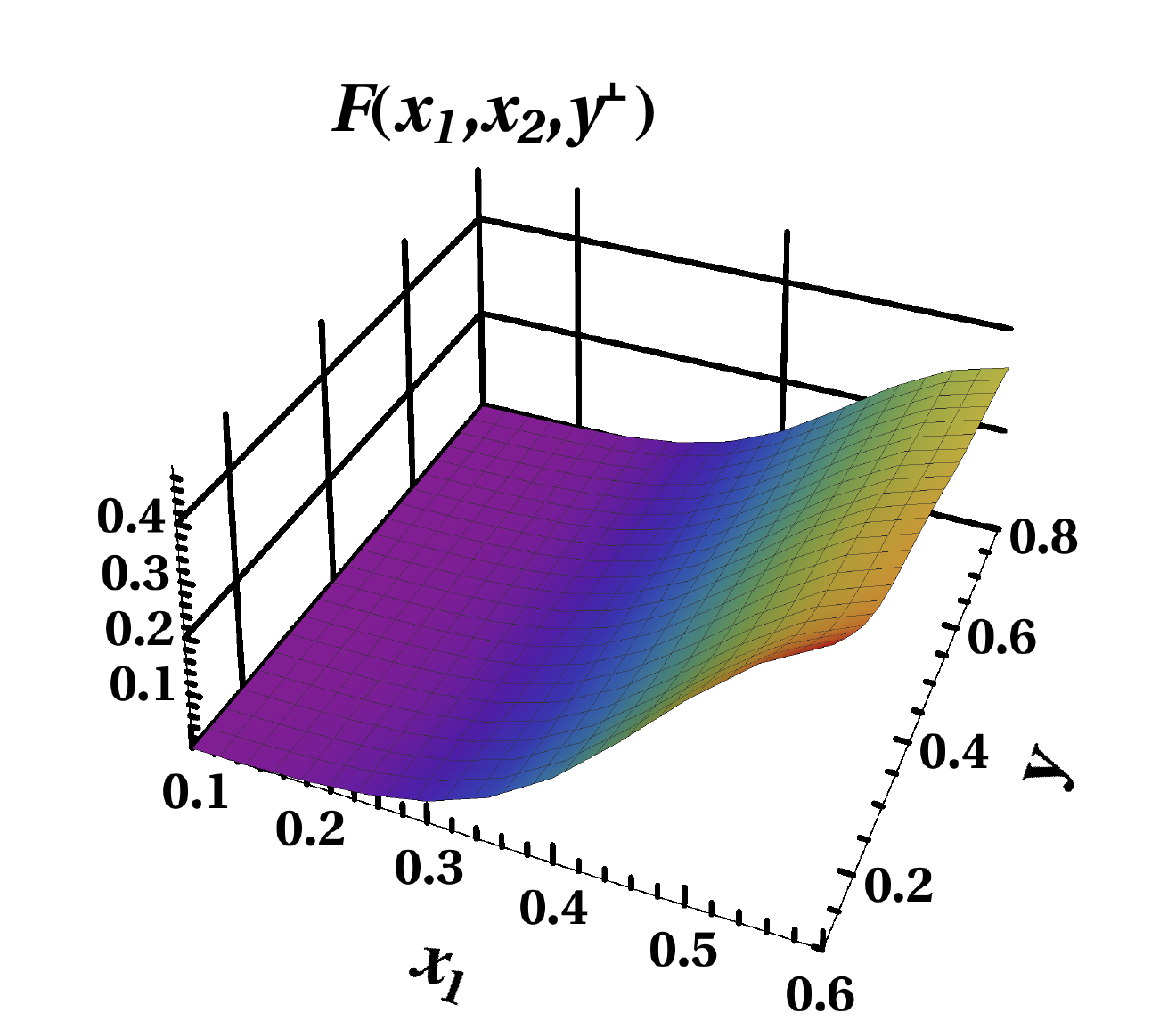}
\caption{(Color online) Plot for $F(x_1,x_2,y^{\perp})$ vs $y^{\perp}$ for fixed value of 
$x_1 = 0.30$ and different values of $x_2 $}\label{fig3}
\end{figure}

Fig~\ref{fig3} shows the 3D plot of the DPD as a function of $y^{\perp}$ and $x_1$ for fixed values of $x_2= 0.3$. The DPD has the maximum but finite value when the relative distance between the $e^-$ and $e^+$ is zero. However, it decreases gradually with increasing value of  $y^{\perp}$ . The magnitude of the distribution increases as the value of $x_1$ increases.

\begin{figure}[!htb]
\centering
\tiny{(a)}\includegraphics[width=8cm,height=6cm,clip]{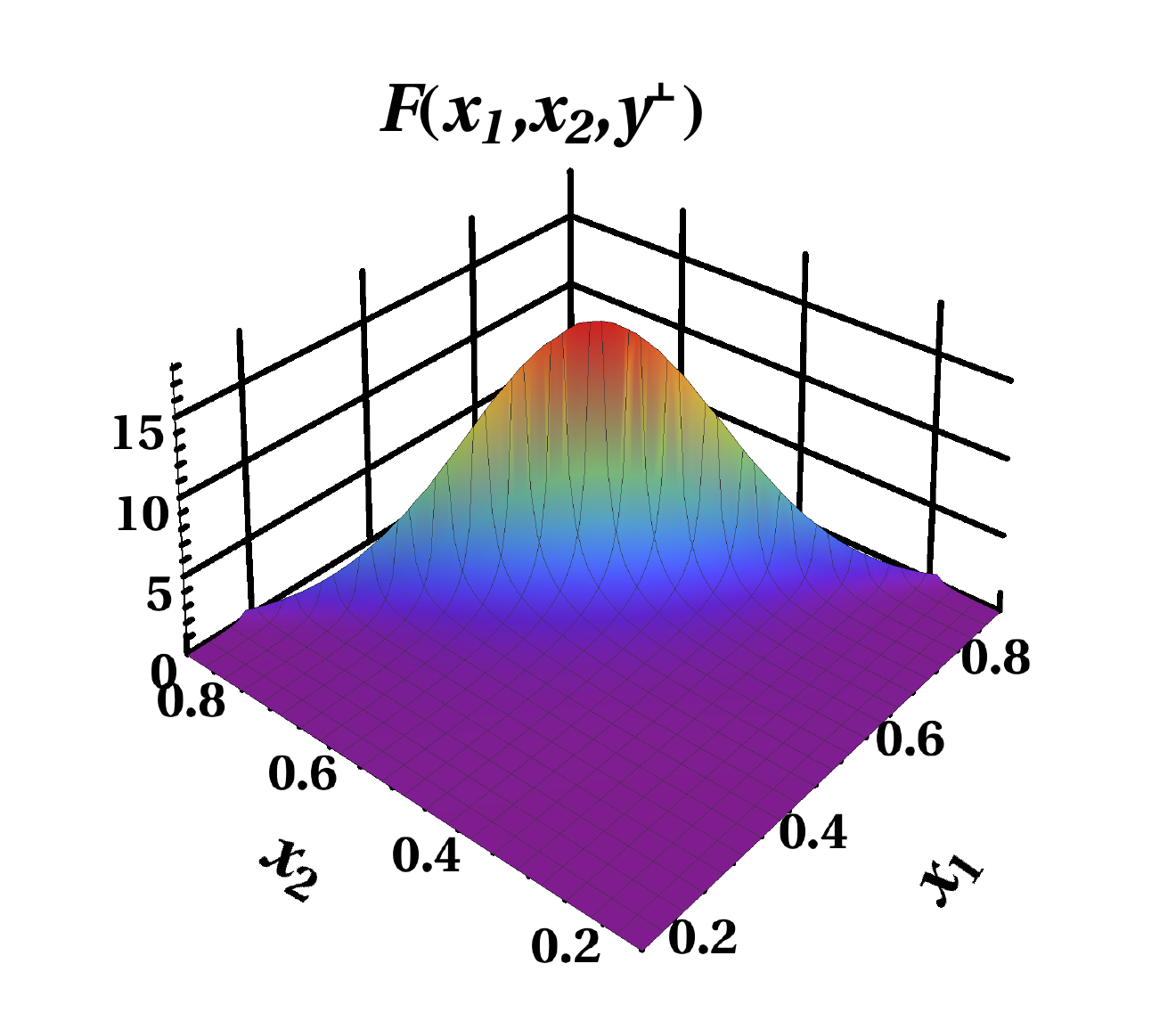}
\tiny{(b)}\includegraphics[width=8cm,height=6cm,clip]{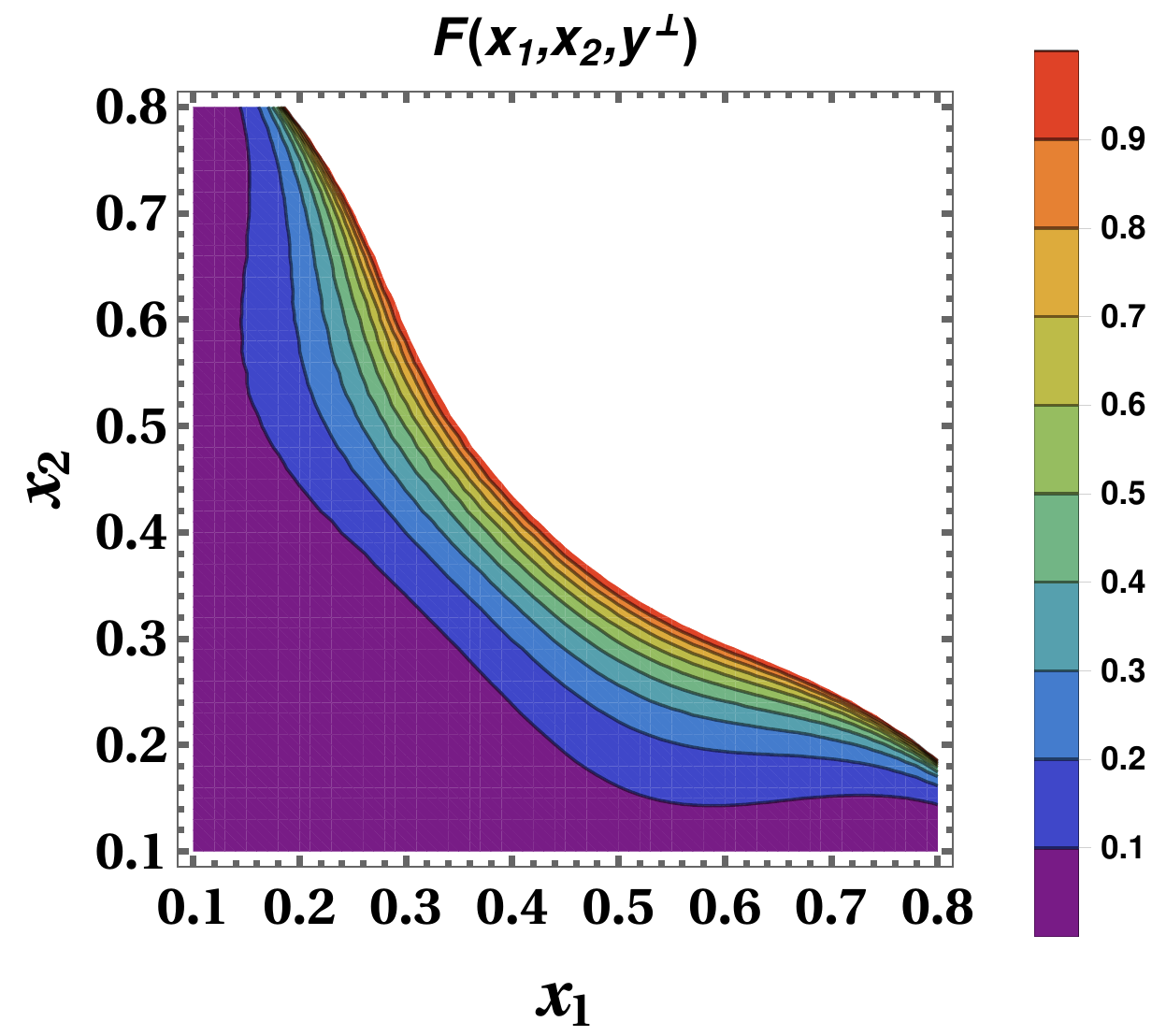}
\caption{ (Color online) 3D (a) and contour (b) plot for the DPD as a function of $x_1$ and $x_2$ for fixed value of $y^{\perp} = 0.2$}\label{fig4}
\end{figure}

Fig~\ref{fig4} (a) shows the 3D plot of the DPD as a function of $x_1$ and $x_2$ for fixed values of $y^{\perp} = 0.2$. The same plot with reduced range for the magnitude of DPD has been shown as a contour plot in Fig~\ref{fig4} (b). Both plots cover the region $x_1 + x_2 <1$. As observed in the 2D plots in Fig~\ref{fig1}, we observe that the magnitude of the DPD sharply increases as $x_1 + x_2 \rightarrow 1$. We can clearly observe the symmetry between $x_1$ and $x_2$ from these two plots.

\begin{figure}[!htb]
\centering
\tiny{(a)}\includegraphics[width=8.0cm,height=6cm,clip]{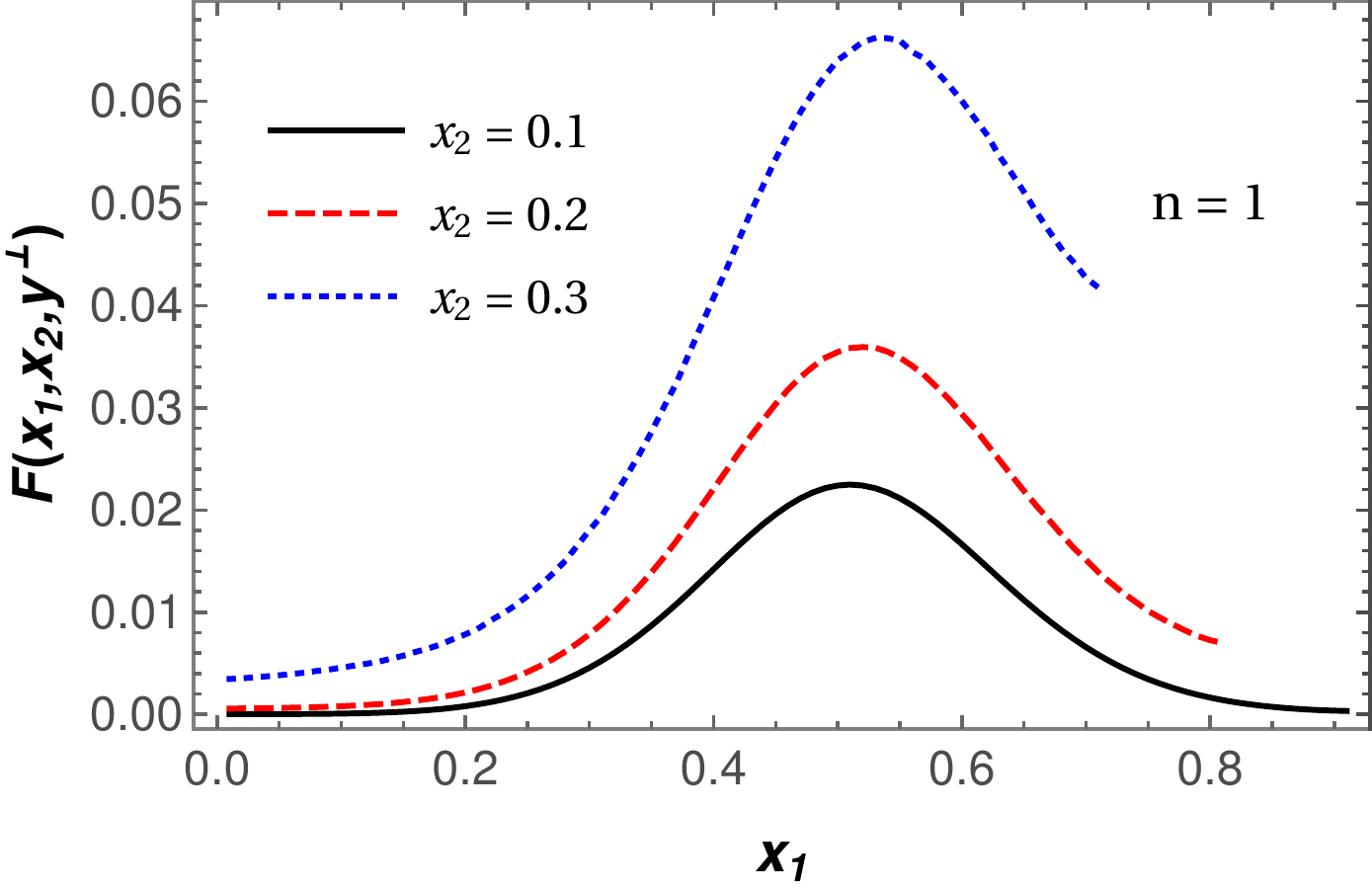}
\tiny{(b)}\includegraphics[width=8.0cm,height=6cm,clip]{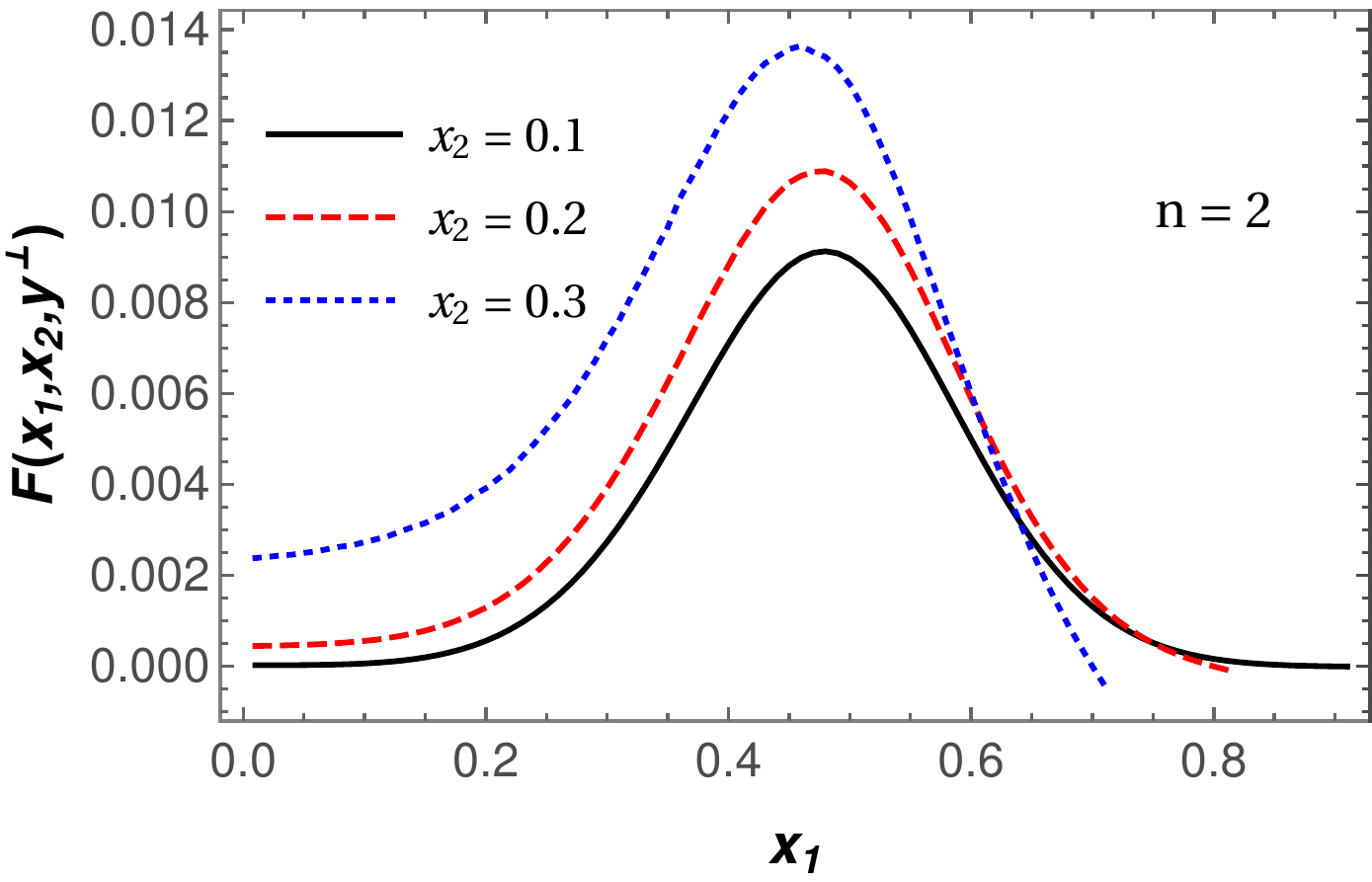}
\tiny{(c)}\includegraphics[width=8.0cm,height=6cm,clip]{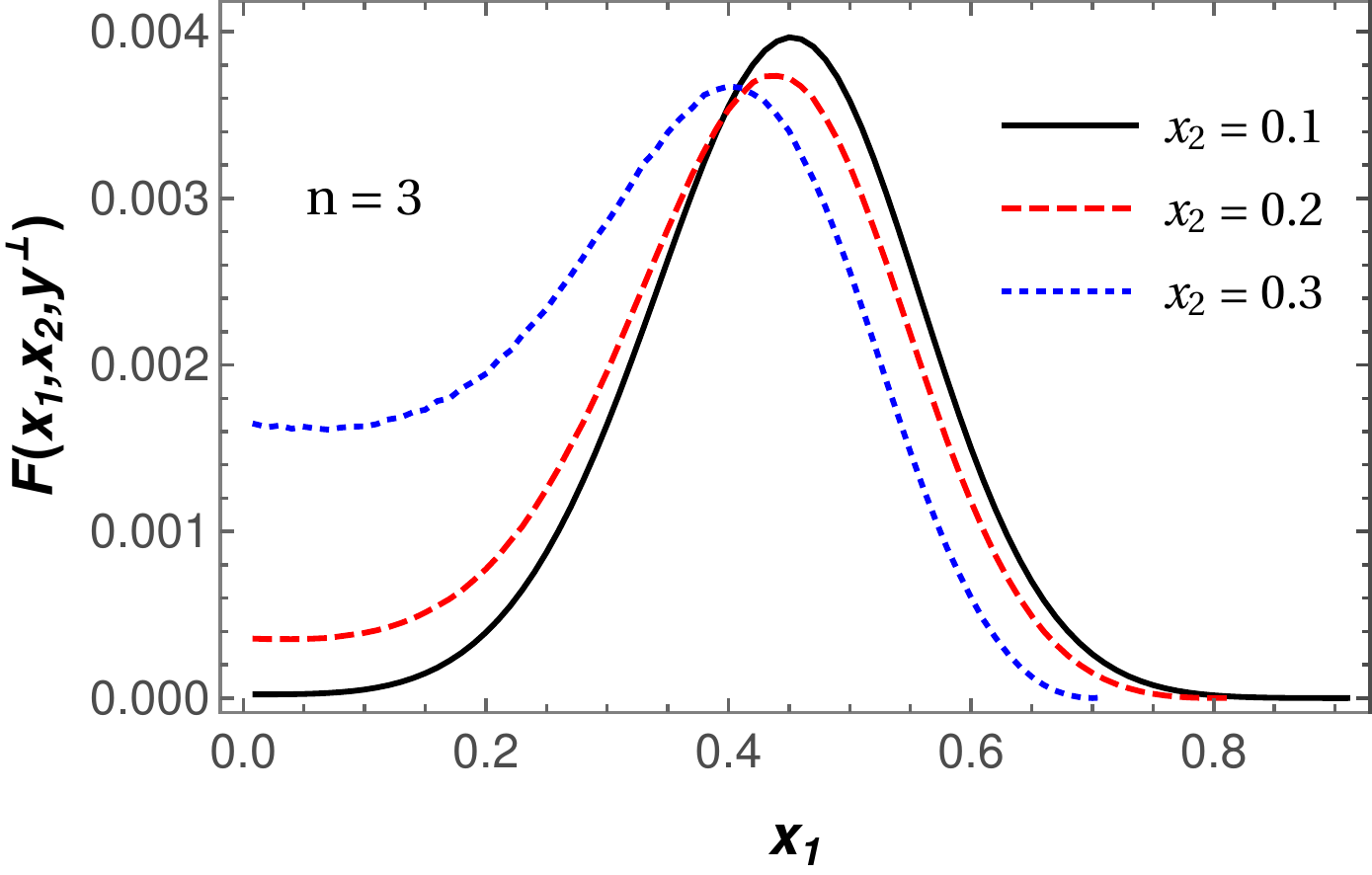}
\caption{(Color online) Plot for $F(x_1,x_2,y^{\perp})$ vs $x_1$ for fixed value of 
$y^{\perp} = 0.2$ and $x_2 = (0.1,0.2,0.3)$. The plots $(a,b,c)$ are for three different value of the parameter $n = (1,2,3)$. }
\label{fig5}
\end{figure}

As discussed in the introduction, the  DPDs for a three-particle system should vanish in the unphysical region $x_1+x_2 > 1$. In some model calculations for example in the Bag model \cite{Mel_19} and in constituent quark model \cite{cqm1} this support property was found to be violated. It order to have a correct behavior near the kinematical bound $x_1+x_2 =1$ it is common to introduce a factor $\theta(1-x_1-x_2) (1-x_1-x_2)^n$, in model calculations of the DPDs for a nucleon,  where $n$ is a parameter to be determined phenomenologically \cite{Koro, Rinaldi:2016jvu}. In \cite{Gaunt:2009re} it was found that   that  a common factor $(1-x_1-x_2)^n$ multiplying all DPDs lead to a violation of momentum sum rule, and the authors suggest a modification of this factor. Motivated by these results,  
In Fig~\ref{fig5} we multiply the DPD by a factor of $(1-x_1-x_2)^n$ and we plot the DPD as a function of $x_1$ for fixed value of $x_2$ and $y^{\perp}$. We show the result for three different values of $n = (1,2,3)$. The behavior near the bound $x_1+x_2 \to 1$ is improved. The DPD has a peak in $x_1$ for fixed value of $x_2 $ (and vice-versa); the position of the peak shifts to lower $x_1$ value as $n$ increases. For a given value of $n$, the peak  occurs at smaller $x_1$ values for larger  $x_2$ .


\section{Conclusion}\label{conclusion}
We have presented a calculation of the electron-positron unpolarized DPD for a positronium-like bound state in light-front QED. We have expressed the DPDs as overlaps of the three-particle LFWFs , that includes a photon. The analytic form of the LFWFs is obtained using LF QED Hamiltonian. Our approach allows us to investigate the correlation between the momentum fractions $x_1, x_2$ and the transverse separation $y_\perp$ of the DPDs without assuming any factorization between them, and may help in improving model parameterizations of nucleon DPDs. The DPDs show strong correlations between these variables. The behavior near the kinematical boundary $x_1+x_2=1$ is improved by introducing a phenomenological factor. Our calculation may act as a guide to develop models for the DPDs of the nucleon at low momentum scale.

\section{ Acknowledgments}
The work of CM is supported by the China Postdoctoral Science Foundation (CPSF) under the Grant No. 2017M623279 and the National Natural Science Foundation of
China (NSFC) under the Grant No. 11850410436. SN is supported by the China Postdoctoral Council under the 
International Postdoctoral Exchange Fellowship Program. We thank T. Kasemets for helpful discussions. 
\\

\section{ Appendix}
Operator for the two fermion unpolarized  correlator:
\be
\mathcal{O}_1(y,z_1)&=&\bar{\psi}(y-\frac{z_1}{2})\gamma^+\psi(y+\frac{z_1}{2})=2\xi^{\dagger}(y-\frac{z_1}{2})\xi(y+\frac{z_1}{2})\nonumber\\
\mathcal{O}_2(0,z_2)&=&\bar{\psi}(\frac{z_2}{2})\gamma^+\psi(-\frac{z_2}{2})=2\xi^{\dagger}(\frac{z_2}{2})\xi(-\frac{z_2}{2})
\ee
with
\be
	\xi(x) &=& \sum_\lambda \chi_\lambda \int {dk^+d^2k^\bot
		\over 2(2\pi)^3 \sqrt{k^+}}\Big(b_\lambda(k)e^{-ikx} + 
		d_{-\lambda}^\dagger(k)e^{ikx} \Big) \, ,
\ee
\be
\mathcal{O}_2(0,z_2)\mathcal{O}_1(y,z_1)&=&\sum_{spin}~\int ~[dk_1]~[dk_1']~[dk_2]~[dk_2']\nonumber\\
&\times&\Big[b^{\dagger}_{\sigma_2}(k_2)b_{{\sigma'}_2}({{k'}}_2)b^{\dagger}_{\sigma_1}(k_1)b_{{\sigma'}_1}({k'}_1)~e^{ik_1.(y-\frac{z_1}{2})}e^{-i{k}_1'.(y+\frac{z_1}{2})}e^{\frac{i}{2}{k}_2'.z_2}e^{\frac{i}{2}{k}_2.z_2}\nonumber\\
&+&b^{\dagger}_{\sigma_2}(k_2)b_{{\sigma'}_2}({{k'}}_2)d_{-\sigma_1}(k_1)d^{\dagger}_{-{\sigma'}_1}({k'}_1)~e^{-ik_1.(y-\frac{z_1}{2})}e^{i{k}_1'.(y+\frac{z_1}{2})}e^{\frac{i}{2}{k}_2'.z_2}e^{\frac{i}{2}{k}_2.z_2}\nonumber\\
&+&b^{\dagger}_{\sigma_2}(k_2)d^{\dagger}_{-{\sigma'}_2}({{k'}}_2)d_{-\sigma_1}(k_1)b_{{\sigma'}_1}({k'}_1)~e^{-ik_1.(y-\frac{z_1}{2})}e^{-i{k}_1'.(y+\frac{z_1}{2})}e^{-\frac{i}{2}{k}_2'.z_2}e^{\frac{i}{2}{k}_2.z_2}\nonumber\\
&+&d_{-\sigma_2}(k_2)b_{{\sigma'}_2}({{k'}}_2)b^{\dagger}_{\sigma_1}(k_1)d^{\dagger}_{-{\sigma'}_1}({k'}_1)~e^{ik_1.(y-\frac{z_1}{2})}e^{i{k}_1'.(y+\frac{z_1}{2})}e^{\frac{i}{2}{k}_2'.z_2}e^{-\frac{i}{2}{k}_2.z_2}\nonumber\\
&+&d_{-\sigma_2}(k_2)d^{\dagger}_{-{\sigma'}_2}({k'}_2)b^{\dagger}_{\sigma_1}(k_1)b_{{\sigma'}_1}({k'}_1)~e^{ik_1.(y-\frac{z_1}{2})}e^{-i{k}_1'.(y+\frac{z_1}{2})}e^{-\frac{i}{2}{k}_2'.z_2}e^{-\frac{i}{2}{k}_2.z_2}\nonumber\\
&+&d_{-\sigma_2}(k_2)d^{\dagger}_{-{\sigma'}_2}({k'}_2)d_{-\sigma_1}(k_1)d^{\dagger}_{-{\sigma'}_1}({k'}_1)~e^{-ik_1.(y-\frac{z_1}{2})}e^{i{k}_1'.(y+\frac{z_1}{2})}e^{-\frac{i}{2}{k}_2'.z_2}e^{-\frac{i}{2}{k}_2.z_2}\Big]\nonumber\\
\ee


\end{document}